%% file: 2011fe_late_Taubenberger.tex
\documentclass[useAMS,usegraphicx,usenatbib]{mn2e}
\usepackage{color}
\usepackage{wasysym}
\usepackage{url}
\usepackage{multirow}
\usepackage[tight]{subfigure}
\usepackage{graphicx}
\usepackage{verbatim}
\usepackage{times}

\def\kms{km\,s$^{-1}$}

\def\OI{O\,{\sc i}}

\def\SiII{Si\,{\sc ii}}

\def\FeI{Fe\,{\sc i}}
\def\FeII{Fe\,{\sc ii}}
\def\FeIII{Fe\,{\sc iii}}

\def\NiII{Ni\,{\sc ii}}

\def\Cofs{$^{56}$Co}

\def\dm15{$\Delta m_{15}(B)$}

\def\lesssim{\mathrel{\hbox{\rlap{\hbox{\lower4pt\hbox{$\sim$}}}\hbox{$<$}}}}
\let\la=\lesssim
\def\gtrsim{\mathrel{\hbox{\rlap{\hbox{\lower4pt\hbox{$\sim$}}}\hbox{$>$}}}}
\let\ga=\gtrsim

\input{journals.tex}

\title[Spectroscopy of SN~2011fe past 1000 days]{Spectroscopy of the Type~Ia supernova 2011fe past 1000 days}
\author[Taubenberger et al.]{S. Taubenberger$^{1,2}$,
N. Elias-Rosa$^{3,4}$, W.\,E.~Kerzendorf$^{5}$, S.~Hachinger$^{6}$, J.~Spyromilio$^{1}$,  
\newauthor  C.~Fransson$^{7}$, M.~Kromer$^{7}$, A.\,J.~Ruiter$^{8,9}$, I.\,R.~Seitenzahl$^{8,9}$, S.~Benetti$^{3}$, E.~Cappellaro$^{3}$, 
\newauthor  A.~Pastorello$^{3}$, M.~Turatto$^{3}$ \& A.~Marchetti$^{10}$\\ 
$^{1}$European Southern Observatory, Karl-Schwarzschild-Str.~2, 85748 Garching, Germany\\
$^{2}$Max-Planck-Institut f\"ur Astrophysik, Karl-Schwarzschild-Str.~1, 85741 Garching, Germany\\
$^{3}$INAF Osservatorio Astronomico di Padova, Vicolo dell'Osservatorio 5, 35122 Padova, Italy\\
$^{4}$Institut de Ci\`encies de l'Espai (CSIC-IEEC), Campus UAB, Torre C5, 2a planta, 08193 Barcelona, Spain\\
$^{5}$Department of Astronomy and Astrophysics, University of Toronto, 50~Saint George Street, Toronto, ON M5S 3H4, Canada\\
$^{6}$Universit\"at W\"urzburg, Lehrstuhl f\"ur Astronomie\,/\,Lehrstuhl f\"ur Mathematik IX, Emil-Fischer-Str.~31\,/\,30, 97074 W\"urzburg, Germany\\
$^{7}$The Oskar Klein Centre, Department of Astronomy, Stockholm University, Albanova, 10691 Stockholm, Sweden\\
$^{8}$Research School of Astronomy and Astrophysics, Mount Stromlo Observatory, Cotter Road, Weston Creek, ACT 2611, Australia\\
$^{9}$ARC Centre of Excellence for All-sky Astrophysics (CAASTRO)\\
$^{10}$INAF - IASF Milano, Via E.~Bassini~15, 20133 Milano, Italy}

\begin{document}

\date{Accepted 2014 December 13. Received 2014 December 10; in original form 2014 November 26}

\pagerange{\pageref{firstpage}--\pageref{lastpage}} \pubyear{2014}

\maketitle

\label{firstpage}

\begin{abstract}
  In this letter we present an optical spectrum of SN~2011fe taken 
  1034\,d after the explosion, several hundred days later than any 
  other spectrum of a Type~Ia supernova (disregarding light-echo 
  spectra and local-group remnants). The spectrum is still dominated 
  by broad emission features, with no trace of a light echo or 
  interaction of the supernova ejecta with surrounding interstellar 
  material. Comparing this extremely late spectrum to an earlier one 
  taken 331\,d after the explosion, we find that the most prominent 
  feature at 331\,d -- [\FeIII] emission around 4700\,\AA\ -- has 
  entirely faded away, suggesting a significant change in the ionisation 
  state. Instead, [\FeII] lines are probably responsible for most of 
  the emission at 1034\,d. An emission feature at 6300--6400\,\AA\ has 
  newly developed at 1034\,d, which we tentatively identify with \FeI\ 
  $\lambda6359$, [\FeI] $\lambda\lambda6231,6394$ or [\OI] $\lambda\lambda6300,6364$. 
  Interestingly, the features in the 1034-d spectrum seem to be collectively 
  redshifted, a phenomenon that we currently have no convincing explanation 
  for. We discuss the implications of our findings for explosion models, but 
  conclude that sophisticated spectral modelling is required for any firm 
  statement.
\end{abstract}

\begin{keywords}
  supernovae: general -- supernovae: individual: SN~2011fe -- line: identification
\end{keywords}

\section{Introduction}
\label{Introduction}

For several years after the explosion, the luminosity of Type~Ia supernovae 
(SNe~Ia) is powered by the decay of radioactive nuclei synthesised in the 
explosion. At the beginning, the ejecta are still optically thick, and the 
radiation is released on photon-diffusion time scales. About 100--200\,d 
later, the ejecta have expanded enough to become transparent for optical 
photons. During the now-commencing nebular phase, the radioactive-heating and 
radiative-cooling rates are similar, making the bolometric luminosity 
evolution a good tracer of the radioactive energy deposition.

Cooling during the nebular phase is mostly accomplished by forbidden-line 
emission: low-lying levels that are still populated at such late epochs 
often have no permitted transition to the ground state, and collisional 
de-excitation is strongly suppressed owing to the low density. The dominant 
coolants in nebular SNe~Ia are iron-group elements, reflecting the composition 
of the inner ejecta. In particular, optical spectra of SNe~Ia around one year 
after the explosion show a characteristic pattern of [\FeII] and [\FeIII] 
lines.

The past decade has led to a wealth of high-quality late-time spectra of 
SNe~Ia. This made it possible to study nucleosynthesis and geometry effects 
in SNe~Ia in unprecedented detail \citep[e.g.][]{kozma2005a,maeda2010c,
maeda2010b,mazzali2011a,blondin2012a,silverman2013a,taubenberger2013a}. 
However, all these studies concentrated on epochs between 200 and 400\,d 
after the explosion. Beyond 400--500\,d, our knowledge on the spectroscopic 
evolution of SNe~Ia is limited. Whenever a SN~Ia was bright enough to 
perform spectroscopy at such late phases, as e.g. in the cases of SNe~1991T 
or 1998bu, it was dominated by a light echo \citep[][respectively]{schmidt1994a,
cappellaro2001a}, prohibiting the study of actual late-time emission from 
the SN ejecta.\footnote{A spectrum of SN~1972E obtained $\sim$700\,d after 
maximum light \citep{kirshner1975a} suffers from low resolution, poor 
signal-to-noise ratio (S/N) and limited wavelength coverage. A spectrum 
of SN~2005cf taken 614\,d after maximum light \citep{wang2009a} turns 
out to be that of an M-type star upon closer inspection.}

The few model calculations that exist for those phases also suffer 
from numerous uncertainties. To capture the relevant nebular physics 
\citep[e.g.][]{mccray1993a,fransson1994a}, non-thermal processes have 
to be accurately modelled, which is sometimes impossible due to missing 
or inaccurate atomic data. Moreover, owing to the low ejecta densities, 
which lead to increased recombination time scales, departures from steady 
state arise. As a consequence, ionisation freeze-out may occur 
\citep{fransson1993a,fransson1996a}. For this reason, a fully 
time-dependent treatment becomes essential after $\sim$500\,d 
\citep{sollerman2004a} to predict the correct ionisation state of the 
ejecta. Spectra observed at epochs $> 500$\,d could greatly help to assess 
the correctness of model calculations, and hence provide a big step 
forward in our understanding of both nebular physics and SN~Ia explosions.

With SN~2011fe \citep{nugent2011a} this goal is now for the first time in 
reach. Its proximity ($d = 6.4$ Mpc; \citealt{shappee2011a}), low dust 
extinction and relatively uncrowded environment make SN~2011fe the ideal 
object to push observations to new limits. 
\citet{kerzendorf2014a} recently reported multi-band optical photometry 
of SN~2011fe between 900 and 950\,d after the explosion, concluding that 
the light-curve decline is consistent with radioactive decay, and that 
there is no evidence for positron escape, an infrared catastrophe (IRC; 
\citealt{axelrod1980a}), dust formation, or a light echo. The derived 
colours were still remarkably blue. Here, we present a spectrum 
of SN~2011fe taken $\sim$100\,d later -- the first nebular spectrum of a 
SN~Ia ever obtained at more than 1000\,d after its explosion -- and compare 
it with a spectrum taken after $\sim$1\,yr.%after the explosion.

\section{Data acquisition and reduction}
\label{Data}

A spectrogram of SN~2011fe was obtained on 2012 July 20.02 (UT dates are used 
throughout this letter), 331 rest-frame days after its inferred explosion on 
2011 August 23.687 \citep{nugent2011a}, with the OSIRIS spectrograph at the 
Gran Telescopio Canarias (GTC). Two grisms (R1000B and R1000R) were used, 
with an exposure time of 300\,s for each grism, and a 1.0-arcsec slit aligned 
along the parallactic angle. Basic CCD reductions and a variance-weighted 
extraction of the spectra were carried out within \textsc{iraf}\footnote{\textsc{iraf} \
is distributed by the National Optical Astronomy Observatory, which is operated 
by the Association of Universities for Research in Astronomy under 
cooperative agreement with the National Science Foundation.}. The 
wavelength calibration was accomplished using arc-lamp exposures and checked 
against night-sky lines. A spectrophotometric standard star observed during 
the same night as the SN was used for flux calibration and telluric-feature 
removal.

\begin{figure}
  \centering 
  \includegraphics[width=0.985\linewidth]{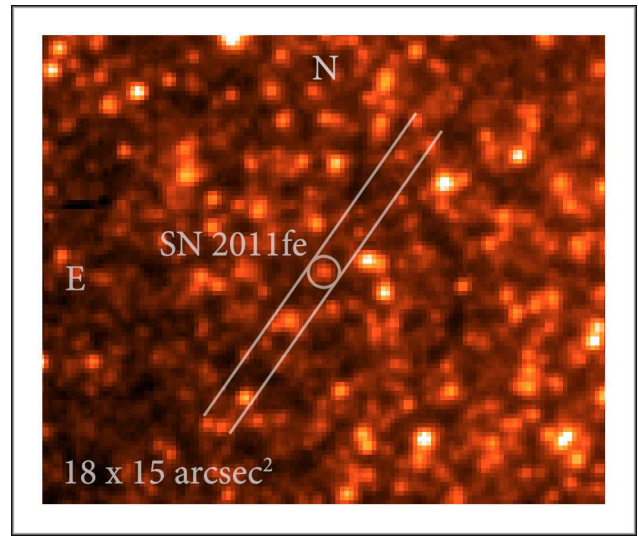}
  \caption{$18\times15$-arcsec$^2$ section of an $i'$-band image, taken 
  with the Gemini-N Telescope + GMOS-N four days after our LBT spectrum. 
  North is up, east to the left. The width and orientation of the 1-arcsec 
  slit used for the LBT observations are indicated.}
  \label{fig:slit}
\end{figure}

SN~2011fe was again targeted on 2014 June 23.22, 1034 rest-frame days after 
the explosion, when the SN had faded to $i' \sim 24$. The observations were 
carried out at the Large Binocular Telescope (LBT), equipped with the MODS1 
dual-beam spectrograph, during Italian\,/\,INAF time. Three exposures of 
3600\,s each were taken in good seeing conditions (0.6 to 1.0 arcsec FWHM) 
through a 1.0-arcsec slit, aligned along the mean parallactic angle over the 
time of the observations (see Fig.~\ref{fig:slit}). A dichroic split the 
light beam at 575\,nm, and the G400L and G670L gratings were used as dispersers 
for the blue and red channel, respectively. The data were pre-reduced using the 
modsCCDRed package\footnote{http://www.astronomy.ohio-state.edu/MODS/Software/modsCCDRed/}, 
and the extraction and calibration of the spectra followed the same scheme as 
described for the GTC data. Eventually, the individual medium-resolution 
spectra were combined and rebinned to 5-\AA\ bins.

\section{Discussion}
\label{Discussion}

\subsection{Line identification}
\label{line identification}

\begin{figure}
  \centering 
  \includegraphics[width=\linewidth]{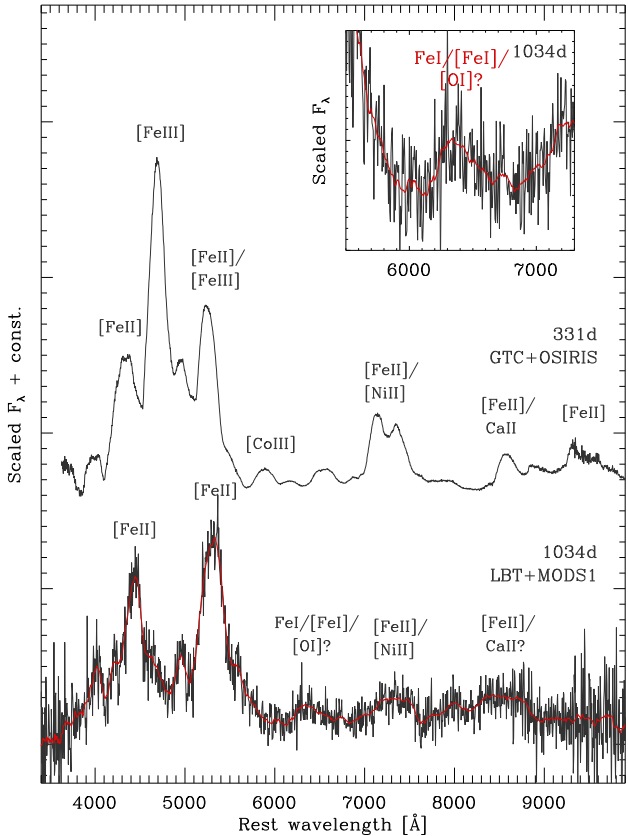}
  \caption{Nebular spectra of SN~2011fe, along with an attempted line 
  identification. A smoothed version (with a boxcar of $\sim$5000 \kms; 
  solid red line) is overlaid on the 1034-d spectrum to facilitate the 
  assessment of line profiles. The inset enlarges the region of the 
  possible \FeI\ or [\OI] feature.}
  \label{fig:lineID}
\end{figure}

In Fig.~\ref{fig:lineID} our nebular spectra of SN 2011fe are presented. 
At +331d the line identification is relatively straightforward and follows 
earlier work in this area (for example, see \citealt{maeda2010c}). The strong 
emissions between 4000 and 5500\,\AA\ are well fitted by non-LTE excitation of 
iron by a $\sim$\,4000\,K thermal electron gas \citep[e.g.][]{axelrod1980a}. 
The feature around 4700\,\AA\ is dominated by [\FeIII], with only a small 
contribution from [\FeII]. The 5250-\AA\ feature is a blend of [\FeIII] and 
[\FeII]. Other features, notably the [\FeII] $\lambda7155$\,/\,[\NiII] 
$\lambda7378$ blend around 7200\,\AA, are marked in Fig.~\ref{fig:lineID}. 

By day 1034 dramatic changes have occurred. The formerly very prominent 
[\FeIII] line at 4700\,\AA\ has disappeared, while the 4400- and 5300-\AA\ 
features have preserved their line ratios and now dominate the flux in the 
optical regime. The 7200-\AA\ feature has weakened relative to the strong 
emission in the 5000-\AA\ region. The comparison of the 1034- and 331-d 
spectra naturally leads to the conclusion that the ionisation structure 
of the ejecta has changed and only little \FeIII\ is present. 

However, the spectrum is difficult to reconcile with thermal excitation. 
The 7200-\AA\ feature has a contribution from \FeII\ multiplet 
14F (a\,$^4$F\,--\,a\,$^2$G) with an upper level at $\sim$2\,eV, whereas 
the bulk of the 4000--5500-\AA\ lines arise from \FeII\ multiplets [e.g. 
6F (a\,$^6$D\,--\,b\,$^4$F), 18F (a\,$^4$F\,--\,b\,$^4$P), 19F 
(a\,$^4$F\,--\,a\,$^4$H)] with upper levels near 2.5\,eV. Only a hot 
($\sim$8000\,K) electron gas could reproduce this spectrum. This is an 
unrealistically high temperature for a SN 1000\,d after explosion, and in 
conflict with the observed low ionisation. 
There are likely contributions by \FeI\ multiplets 2F (a\,$^5$D\,--\,a\,$^5$P) 
and 3F (a\,$^5$D\,--\,a\,$^3$P2) to the 5300-\AA\ feature, and by multiplets 
4F (a\,$^5$D\,--\,b\,$^3$F2) and 6F (a\,$^5$D\,--\,b\,$^3$P) to the 4400-\AA\ 
feature. Again, the observed ratio of the 4400- to the 5300-\AA\ feature 
is not in agreement with thermal excitation at any realistic temperature. 
Recombination or non-thermal excitation processes must therefore prevail.

The series of cobalt lines between 5800 and 6700\,\AA, still fairly prominent 
in the 331-d spectrum, is no longer detected at 1034\,d. Given that most of 
the cobalt in SNe~Ia is \Cofs, which decays with a half-life of 77\,d, the 
observed fading of the cobalt lines is expected. However, in the same region 
a new, broad emission feature has emerged (see inset of Fig.~\ref{fig:lineID}). 
It is centred around 6360\,\AA, and might be identified with some combination 
of \FeI\ $\lambda6359$, [\FeI] $\lambda\lambda6231,6394$ and [\OI] 
$\lambda\lambda6300,6364$. Given the dominance of [\FeII] lines in the spectrum 
and our proposal that [\FeI] contributes in the 4000--5500-\AA\ range, the 
identification with \FeI\ lines may appear more natural. On the other hand, if 
[\OI] can be confirmed through spectral modelling, this would be the third 
detection of [\OI] $\lambda\lambda6300,6364$ in the nebular spectrum of a 
thermonuclear SN, after SN~1937C \citep{minkowski1939a} and SN~2010lp 
\citep{taubenberger2013b}, though this time only at a much later epoch. 

There is no sign of narrow or intermediate-width emission lines that might hint 
at interaction with interstellar material (ISM). The transition of SN~2011fe 
into the remnant phase, when the emission becomes dominated by the shock 
originating from ejecta-ISM collisions, has not yet started. In particular, 
no narrow H$\alpha$ line is detected. Such a line might be expected in nebular 
spectra of single-degenerate explosions, where H-rich material is stripped 
from a non-degenerate companion star upon the impact of the SN ejecta 
\citep{marietta2000a,pakmor2008a,liu2012a}. Thus far, H$\alpha$ arising from 
stripped material has never been detected in nebular SN~Ia spectra 
\citep{mattila2005a,leonard2007a}. With our 1034-d spectrum, we now extend 
the series of non-detections to much later epochs. Though the presence of a 
very weak H$\alpha$ line cannot be excluded because of S/N limitations, our 
non-detection is in line with the absence of narrow H$\alpha$ in our 331-d 
spectrum and in an even higher-S/N spectrum of SN~2011fe taken 275\,d after 
$B$-band maximum \citep{shappee2013a}.

\citet{kerzendorf2014a} discussed the possibility of a light echo, as in their 
photometry taken about 950\,d after the explosion SN~2011fe appeared bright 
and blue. They argued, however, that a strong light echo -- as  e.g. observed 
in SNe~1991T and 1998bu \citep[][respectively]{schmidt1994a,cappellaro2001a} -- 
was unlikely, since the observed colours did not agree with those of SNe~Ia 
around maximum light, as one would expect for a light echo. With our 1034-d 
spectrum we can now finally rule out the possibility of a light echo, since no 
\mbox{(pseudo-)\,continuum} or P-Cygni features are detected.

\subsection{Line shifts}
\label{Line shifts}

Once a SN has turned transparent to optical photons, emission-line 
profiles probe the underlying emissivity distribution in the ejecta. The 
latter depends on the relative location of coolants and radioactive material, 
and on the mean free paths of $\gamma$-rays and positrons. During the epochs 
under consideration in this letter ($>300$\,d after the explosion), the 
ejecta are largely transparent to $\gamma$-rays, and the energy deposition 
is dominated by positrons and electrons \citep{milne2001a,seitenzahl2009d}. 
Studies of the late-time bolometric luminosity of SNe~Ia \citep{cappellaro1997b,
leloudas2009a,kerzendorf2014a} have argued for almost complete positron 
trapping even as late as 900\,d, suggesting a rather short mean free path of 
positrons, and a nearly in-situ deposition of the radioactive-decay energy 
during the positron-dominated phase. 

Detailed studies of nebular optical and infrared emission-line profiles in 
SNe~Ia have been carried out in the past \citep{mazzali1998a,motohara2006a,
gerardy2007a,maeda2010c,maeda2010b}. An interesting result of such studies 
was that certain emission lines (e.g. [\FeII] $\lambda7155$ and [\NiII] 
$\lambda7378$; \citealt{maeda2010c}) showed significant blue- or redshifts 
in different SNe, which was interpreted as a viewing-angle effect. 
\citet{maeda2010b}, finally, found a correlation between the post-maximum 
velocity gradient in \SiII\ $\lambda6355$ \citep{benetti2005a} and the shift 
of nebular [\FeII] $\lambda7155$ and [\NiII] $\lambda7378$ lines. In our 
331\,d spectrum of SN~2011fe we find both lines to be blueshifted, the 
[\FeII] $\lambda7155$ line by almost 1000 \kms, the [\NiII] $\lambda7378$ 
line by 1100 \kms. This meets the expectations, given that SN~2011fe is a 
low-velocity-gradient SN in the \citet{benetti2005a} classification scheme 
\citep[e.g.][]{pereira2013a}.

\begin{figure*}
  \centering 
  \includegraphics[width=0.88\linewidth]{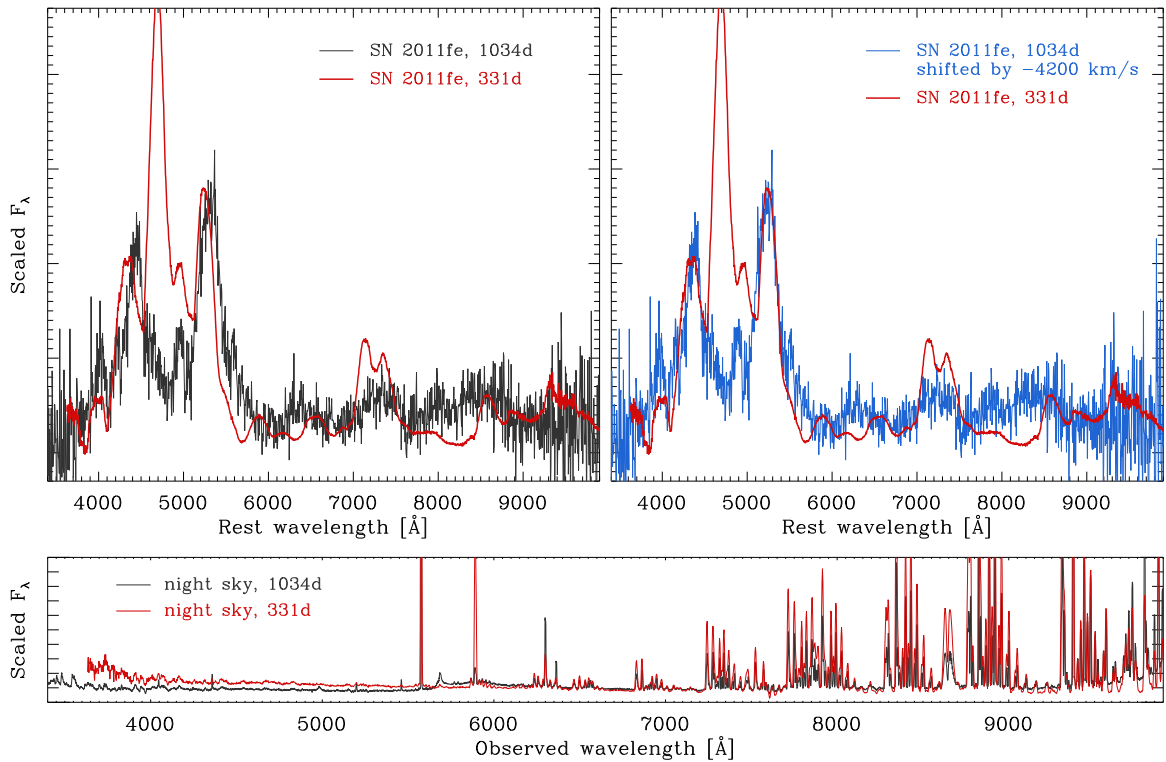}
  \caption{\textit{Top left panel:} Superposition of the 1034-d and 331-d 
  spectra of SN~2011fe, scaled arbitrarily in flux. A wavelength offset 
  seems to be present in all features that can be identified in both spectra. 
  \textit{Top right panel:} The match of features is substantially improved 
  if a wavelength shift corresponding to $-4200$ \kms\ is applied.
  \textit{Bottom panel:} No offset is present in the night-sky spectra.}
  \label{fig:shift}
\end{figure*}

In the 1034-d spectrum the S/N is insufficient to directly measure accurate 
positions for all except the two strongest emission lines. However, as noted 
in the previous section, the 331-d and 1034-d spectra appear to have several 
[\FeII] features in common, and so a superposition of the two spectra 
(Fig.~\ref{fig:shift}, top panels) can be instructive to determine changes 
in the line profiles or positions. From Fig.~\ref{fig:shift} it is immediately 
evident that there seems to be a global offset in several (if not all) common 
features, in the sense that the 1034-d spectrum appears globally redshifted 
with respect to the 331-d spectrum. The shift amounts to 4200 \kms, as 
inferred from a cross-correlation of the peaks in the 4000--5500\,\AA\ region, 
with an estimated uncertainty of $\pm 200$ \kms. That this shift is real is 
demonstrated in the bottom panel of Fig.~\ref{fig:shift}, where no offset 
is seen in the night-sky spectra of the two observations.

Following the paradigm that nebular spectra probe the ejecta geometry, the 
observed shift might be understood if the emitting regions of the ejecta were 
different at 331 and 1034\,d. At 331\,d the energy deposition is dominated 
by positrons from \Cofs\ decay, at 1034\,d by electrons and X-rays from 
$^{57}$Co decay \citep{seitenzahl2009d,roepke2012a}. A spatial separation of 
\Cofs\ and $^{57}$Co could thus lead to the observed effect. However, explosion 
models \citep[e.g.][]{seitenzahl2013a} suggest that \Cofs\ and $^{57}$Co are 
synthesised co-spatially, making this explanation unlikely. The same is true 
for the attempt to explain the observed time evolution by residual opacity 
in the core of the ejecta at 331\,d. First, a sufficiently large optical depth 
to produce line shifts of 4000 \kms\ almost one year after the explosion is 
unreasonable. Second, the lines in the 331-d spectrum are close to their 
rest-frame position, so that we do not have to explain a blueshift at 331\,d, 
but a redshift at 1034\,d.

Since the affected features are blends (actually often multiplets), changes 
in the ionisation and excitation conditions might lead to a strengthening or 
weakening of individual constituents, resulting in an effective shift of the 
entire blend. Also, if there is indeed a significant contribution by lines 
from \FeI\ at 1034\,d, wavelength shifts compared to the 331-d spectrum are 
to be expected. However, it appears rather unlikely that any of these effects 
leads to the same offset in \textit{all} features throughout the spectrum, 
though ultimately this has to be verified by accurate spectral modelling.

\subsection{Implications for models}
\label{Implications for models}

Comparing the 1034-d spectrum of SN~2011fe to synthetic nebular spectra 
would be extremely worthwhile to infer details about the composition, 
ionisation and excitation state of the ejecta. Unfortunately, not many model 
calculations with at least broad-band spectral information have ever been 
performed for SNe~Ia at such a late phase. One of the most extended 
non-grey time-dependent model calculations for SNe~Ia to date is that by 
\citet{leloudas2009a}. Their synthetic $U\!BV\!RIJHK$-band light curves of the 
W7 model \citep{nomoto1984a} show a rapid decline starting at $\sim$500\,d, 
which can be attributed to the onset of an IRC \citep{axelrod1980a}. Such an 
IRC occurs when the temperature drops below $\sim$1500\,K, too low to 
populate the upper levels of typical nebular emission lines in the optical 
and near-IR regime. As a consequence, most of the cooling henceforth happens 
via fine-structure lines in the mid- and far-IR, leading to a dramatic change 
of the spectral energy distribution.
 
\citet{kerzendorf2014a} argued that their 930-d photometry of SN~2011fe 
showed no evidence of the enhanced fading predicted by an IRC. This conclusion 
seems to be supported by our 1034-d spectrum, where we still observe prominent 
emission lines in the optical regime. However, there may be ways to reconcile 
our observations with an IRC, if the latter is restricted to certain regions 
of the ejecta while others remain hot\footnote{The critical temperature for an 
IRC is density-dependent; local density enhancements in form of clumping may 
significantly postpone the IRC.}, or if the observed lines are actually 
recombination lines and hence not excited thermally.

As already mentioned in Section~\ref{line identification}, the 1034-d 
spectrum of SN~2011fe shows a broad (FWHM\,$\sim$12\,000 \kms) emission 
feature centred at $\sim$6360\,\AA, for which \FeI\ $\lambda6359$, [\FeI] 
$\lambda\lambda6231,6394$ and [\OI] $\lambda\lambda6300,6364$ are possible 
identifications. Interpreted as [\OI], the feature would be redshifted by 
$\sim$2000 \kms. If this identification is correct, it has important 
consequences for the preferred explosion scenario for SN~2011fe in particular 
and -- given the conception of SN~2011fe as a perfectly `normal' SN~Ia -- for 
the entire SN~Ia class. To produce late-time [\OI] emission with the given 
line profile, oxygen has to be present in the inner part of the ejecta (the 
line profile is not flat-topped, which disfavours emission from a shell), 
which is fulfilled only for a small subset of SN~Ia explosion models 
\citep{taubenberger2013b}, notably violent mergers \citep{pakmor2012a,kromer2013b}. 
However, \citet{jerkstrand2011a} found that at very late phases, when thermal 
collisional excitation of [\OI] $\lambda\lambda6300,6364$ was no longer 
possible, even in the arguably much more oxygen-rich SN~1987A the emission 
feature near 6300\,\AA\ was dominated by \FeI\ recombination lines rather 
than [\OI]. Whether these conditions are met in SN~2011fe at 1034\,d, where 
we still observe relatively high-excitation [\FeII] lines, has to be verified 
by detailed simulations of the plasma state.

\section{Conclusions}
\label{Conclusions}

We have presented two optical spectra of SN~2011fe taken 331 and 1034\,d 
after explosion. At 1034\,d the emission still comes from 
the nebular SN ejecta, with no signs of a light echo or interaction with 
interstellar material. Nonetheless, strong changes have occurred compared 
to the early nebular phase. The most striking of these is the complete 
fading of the 4700-\AA\ [\FeIII] emission -- the by far most prominent 
feature at 331\,d -- which we attribute to a decrease of the ionisation 
state. [\FeII] features, on the contrary, can still be identified. A weak, 
broad emission feature is now present at $\sim$6360\,\AA, which might either 
be attributed to lines from \FeI, or to [\OI] $\lambda\lambda6300,6364$, 
which would have important consequences for explosion scenarios.

From the features that the 331-d and 1034-d spectra seem to have in common 
(mostly [\FeII] blends) a relative wavelength shift can be derived, in the 
sense that at 1034\,d all features appear to be redshifted by $\sim$4000 \kms\ 
compared to the earlier epoch. We currently have no convincing explanation 
for this unexpected behaviour, and it remains unclear whether the origin of 
this shift is geometric, optical-depth-related, or a conspiracy of atomic 
physics.

The detection of prominent emission lines in the 4000--5000\,\AA\ 
range, combined with the late-time luminosity of SN~2011fe reported by 
\citet{kerzendorf2014a}, seems to disfavour the idea that an IRC has taken 
place in the bulk of the ejecta. In contrast, model calculations predict 
the onset of an IRC with strong observable consequences already at 500\,d 
\citep{leloudas2009a}. Whether this discrepancy hints at an inadequacy of 
the W7 explosion model used for those calculations, or at shortcomings in 
the atomic data and nebular-physics treatment, has to be tested in future 
modelling efforts. The 1034-d spectrum of SN~2011fe presented here provides 
the perfect benchmark for such modelling.

\section*{Acknowledgements}

  Observations were carried out using the Gran Telescopio Canarias (GTC), 
  installed in the Spanish Observatorio del Roque de los Muchachos, on the 
  island of La Palma, and the Large Binocular Telescope (LBT) at Mt. Graham, 
  AZ. The authors are grateful to the LBT-Italy consortium for making these 
  ground-breaking DDT observations possible, and would like to thank the 
  telescope operators at LBT, and the support astronomers at GTC, for their 
  commitment. We are also grateful to our referee, Jeffrey Silverman, for 
  his careful reading of the manuscript and his helpful comments.
  
  ST is supported by the Transregional Collaborative Research Centre TRR 33 
  `The Dark Universe' of the DFG. NER acknowledges support from the European 
  Union Seventh Framework Programme (FP7/2007-2013) under grant agreement 
  n. 267251 (AstroFIt). SH is supported by  an ARCHES award. Parts of this 
  research were conducted by the Australian Research Council Centre of 
  Excellence for All-sky Astrophysics (CAASTRO), through project n. 
  CE110001020, and by ARC Laureate Grant FL0992131. SB, EC, AP and MT are 
  partially supported by PRIN-INAF 2011 with the project `Transient Universe: 
  from ESO Large to PESSTO'.

\footnotesize{
  \bibliographystyle{mn2e}
%  \bibliography{astrofritz}

}

\label{lastpage}

\end{document}

%% file: journals.tex
\def\aj{AJ}%
\def\araa{ARA\&A}%
\def\apj{ApJ}%
\def\apjl{ApJ}%
\def\apjs{ApJS}%
\def\aap{A\&A}%
\def\mnras{MNRAS}%
\def\nat{Nature}%